\documentclass[twocolumn,nofootinbib]{revtex4}

\usepackage{amssymb}
\usepackage{amsfonts}
\usepackage{amsmath}
\usepackage[latin1]{inputenc}
\usepackage{graphicx,color}

\makeatletter

\newcommand{\al}{\alpha}

\newcommand{\be}{\begin{equation}}
\newcommand{\ee}{\end{equation}}
\newcommand{\bea}{\begin{eqnarray}}
\newcommand{\eea}{\end{eqnarray}}

\usepackage{young}
\usepackage[vcentermath]{youngtab}

\begin{document}

\title{Sources for Generalized Gauge Fields}
\author{Claudio Bunster$^{1,2,3}$ and Marc Henneaux$^{1,2,4}$}
\affiliation{${}^1$Max-Planck-Institut f\"ur Gravitationsphysik (Albert-Einstein-Institut),
M\"uhlenberg 1, D-14476 Potsdam, Germany}
\affiliation{${}^2$Centro de Estudios Cient\'{\i}ficos (CECs), Casilla 1469, Valdivia, Chile}
\affiliation{${}^3$Universidad Andr\'es Bello, Av. Rep\'ublica 440, Santiago, Chile} 
\affiliation{${}^4$Universit\'e Libre de Bruxelles and International Solvay Institutes, ULB-Campus Plaine CP231, B-1050 Brussels, Belgium}

\begin{abstract}
Generalized gauge fields are tensor fields with mixed symmetries. For gravity and higher spins in dimensions greater than four, the fundamental field in the ``magnetic representation" is a generalized gauge field. It is shown that the analog of a point source for a generalized gauge field is a special type of brane whose worldsheet has another brane interwoven into it: a current within a current.  In the case of gravity in higher dimensions, this combined extended object is the generalization of a magnetic pole.  The product of the  ``electric" and ``magnetic" strengths is quantized.

\end{abstract}
\maketitle

\section{Introduction}
\setcounter{equation}{0}

A generalized gauge field is a tensor with mixed symmetry properties \cite{Curtright:1980yk}. Its indices are divided in groups. It is antisymmetric under exchanges of indices within each group, and there are restrictions under exchanges of indices belonging to different groups.  Generalized gauge fields are automatically brought in if one studies gravitational electric-magnetic duality in space-time dimensions greater than four.  The magnetic dual of a symmetric tensor gauge fields of rank two is then a generalized gauge field \cite{Curtright:1980yk,West:2001as,Hull:2001iu,Boulanger:2003vs}.

The sources of purely antisymmetric gauge fields are extended objects \cite{Kalb:1974yc,Teitelboim:1985ya} whereas those of totally symmetric fields are point particles.  It is shown in this article that the  source for a generalized gauge field is the  worldsheet of a special brane that has another brane interwoven into it: a current within a current.  In the case of linearized gravity in higher dimensions, this combined extended object is the generalization of a magnetic pole.  The quantization condition for product of the ``electric" and ``magnetic" strengths holds.

A self-contained treatment of generalized gauge fields and gravitational electric-magnetic duality, as well as references to previous work on the subject may be found in \cite{Bunster:2013oaa}.  We will adhere to the conventions of that reference.

The plan of the paper is the following: Section \ref{SecII} is devoted to the simplest case, namely the ``magnetic representation" of the linearized gravitational field in $D=5$ space-time dimensions. In Section \ref{General} the analysis is extended to the case of a  generalized gauge field with a two-column Young tableau, which contains, in particular,  the dual description of gravitation in any spacetime dimension.  A brief Section \ref{Quantization}  then establishes the quantization condition for the product of the electric and magnetic strengths.   Finally, Section  \ref{Concluding} is devoted to concluding remarks.

\section{Source  in $D=5$: a current within a current}
\label{SecII}
\setcounter{equation}{0}

The simplest case, which already contains many of the key elements, is the linearized gravitational field in five spacetime dimensions.  The ``electric" representation is, as always, the standard one in terms of the symmetric tensor $h_{\al \beta}$,
\be 
h_{[\al \beta]}= 0.
\ee
The field equation with an electric source is
\be
G^{\al \beta}[h] = T^{\al \beta}  \label{LinEinsEl}
\ee
where $G^{\al \beta}[h]$ is the linearized ``electric" Einstein tensor constructed out of $h_{\mu \nu}$ and its derivatives.  It obeys the symmetry conditions
\be 
G^{[\al \beta]}= 0 \label{SymEl}
\ee
and the identity
\be
\partial_\al G^{\al \beta} = 0.   \label{ConsEl}
\ee
As a consequence of (\ref{LinEinsEl}), these conditions must also be obeyed by the source $T^{\al \beta}$,
\be 
T^{[\al \beta]}= 0 \label{SymTEl}
\ee
\be
\partial_\al T^{\al \beta} = 0.   \label{ConsTEl}
\ee

The energy-momentum tensor of a point source may be written as
\be
T^{\alpha \beta}(x) =  \int dz^{\alpha} \delta^{(D)}(x-z(\xi)) P^\beta (\xi)   .  \label{EMTEl}
\ee
Here,  $x = z(\xi)$ is the  equation of the particle's worldline in terms of an arbitrary parametrization. In order for the integral (\ref{EMTEl}) to be reparametrization invariant, $P^\beta(\xi)$ must be a scalar in $\xi$.   The conservation law (\ref{ConsTEl}) holds if and only if 
\be
 \frac{d P^\beta}{d \xi} = 0, \label{ConsP0}
\ee
whereas the symmetry property (\ref{SymTEl}) implies 
\be 
P^{\beta} = m \frac{dz^\beta}{d \tau}  \label{ExprP0}
\ee
where $\tau$ is the proper time and $m$ the ``electric" mass.  This equation, combined with (\ref{ConsP0}), implies that both $m$ and $\frac{dz^\beta}{d \tau}$ do not change along the world line, that is 
\begin{eqnarray}
&&\frac{dm}{d \tau} = 0 \\
&& \frac{d^2 z^\beta}{d \tau^2} = 0.
\end{eqnarray}
The fact that the conservation law of the source restricts its worldline, i.e., it gives its equations of motion, is intimately connected with the presence of the spacetime index $\beta$ in the strength $P^\beta$ of the source.  It does not arise for pure $p$-forms where the corresponding strength is a spacetime scalar.

In the magnetic representation, the counterpart of the symmetric $h_{\al \beta}$ is a generalized gauge field $t_{\alpha_1 \alpha_2 \beta}$ with the symmetry properties of the Young tableau
$$
\yng(2,1)
$$
i.e., 
\be
t_{[\alpha_1 \alpha_2] \beta} = t_{\alpha_1 \alpha_2 \beta}, \; \; \; \; t_{[\alpha_1 \alpha_2 \beta]} = 0.
\ee
This field was denoted by a capital $T$ in \cite{Bunster:2013oaa}, we reserve here that letter for the corresponding energy-momentum tensor.

The magnetic counterparts of (\ref{LinEinsEl}), (\ref{SymEl}) and (\ref{ConsEl}) are
\be
G^{\al_1 \al_2 \beta} = T^{\al_1 \al_2 \beta}_{\textrm{mag}}, \label{LinEinsMag}
\ee
with
\be
G^{[\alpha_1 \alpha_2] \beta} = G^{\alpha_1 \alpha_2 \beta}, \; \; \; \; G^{[\alpha_1 \alpha_2 \beta]} = 0, \label{SymMag}
\ee
and
\be
\partial_{\alpha_1} G^{\alpha_1 \alpha_2 \beta} = 0,   \label{ConsMag}
\ee
which implies
$$
\partial_{\beta} G^{\alpha_1 \alpha_2 \beta} = 0.
$$
In the sequel, we shall drop the subscript ``mag"  when no confusion can arise due to the number of indices present. 

As a consequence of (\ref{LinEinsMag}), the conditions (\ref{SymMag}) and (\ref{ConsMag}) must also be obeyed by the source $T^{\alpha_1 \alpha_2 \beta}$,
\be
T^{[\alpha_1 \alpha_2] \beta} = T^{\alpha_1 \alpha_2 \beta}, \; \; \; \; T^{[\alpha_1 \alpha_2 \beta]} = 0 \label{SymEMMag}
\ee
\be
\partial_{\alpha_1} T^{\alpha_1 \alpha_2 \beta} = 0,   \label{ConsEMMag}
\ee

The question is then: what is the magnetic analog of (\ref{EMTEl})?

The answer is suggested by the symmetry properties.  The antisymmetry in $(\al_1, \al_2)$ indicates that a two-dimensional worldsheet should come in. On the other hand, the lone index $\beta$ is naturally associated to the tangent to a worldline.  Furthermore, the vanishing of the totally antisymmetric part signals that the worldline should be interwoven into the worldsheet so that the three-volume spanned by the three tangents (two for the worldsheet, one for the worldline) vanishes.  

We propose then
\be
T^{\alpha_1 \al_2 \beta}(x) =  \int dz^{\alpha_1} \wedge dz^{\al_2} \delta^{(5)}(x-z(\xi))  P^\beta_{\textrm{mag}} (\xi)  \label{EMTMag}
\ee
where $z^{\al} = z^{\al}(\xi^a)$ are the equations of the worldsheet in an arbitrary parametrization $(\xi^a) = (\xi^0, \xi^1)$. 
This magnetic energy-momentum tensor satisfies the  symmetry properties (\ref{SymEMMag}) and the conservation law (\ref{ConsEMMag})
provided: (i) the worldsheet is tangent everywhere to the vector $P^\beta_{\textrm{mag}}$,
\be
P^\beta_{\textrm{mag}} = P^a_{\textrm{mag}} \frac{\partial z^\beta}{\partial \xi^a} \label{ParaP};
\ee
(ii) \be
\frac{\partial P^\beta_{\textrm{mag}}}{\partial \xi^a} = 0; \label{ParTran}
\ee
and, (iii) the worldsheet is
 infinite in all directions or closed (no boundary).

Just as for the particle case, Eqs. (\ref{ParaP}) and (\ref{ParTran}) severely restrict the shape of the worldsheet.  To see how these restrictions come about it is useful to choose the time lines on the worldsheet so that 
\be
P^\beta_{\textrm{mag}} = m_{\textrm{mag}}  \frac{\partial z^\beta}{\partial \tau},  \label{PP2}
\ee
where $\tau$ is the proper time along the lines of constant $\xi^1$.  Then (\ref{ParTran})  implies 
\be
\frac{\partial m_{\textrm{mag}}}{\partial \xi^a} = 0,
\ee
and
\be
z^\beta (\tau, \xi^1) = Z^\beta(\tau) + y^\beta(\xi^1)  \label{Sol0}
\ee
with
\be
\frac{d^2 Z^\beta}{d \tau^2} = 0
\ee  
and $y^\beta (\xi^1)$ arbitrary.  

So we see that the dynamics of the string that sweeps the worldsheet is indeed severely restricted,  much more than for the standard string. Only the ``zero mode" $Z^\beta(\tau)$ is dynamical, and behaves as a free particle.  The string may have an arbitrary initial shape, but once that shape is given, it moves rigidly in space during the course of time, with the spacetime velocity of the zero mode. In more geometrical terms, the worldsheet is obtained by displacing parallelly to itself a straight timelike worldline tangent to $P^\beta_{\textrm{mag}}$ along an arbitrary spatial path.  The worldsheet is thus a ruled surface with parallel generatrices (the straight timelike worldlines tangent  to $P^\beta_{\textrm{mag}}$) and arbitrary directrix (the initial shape $y^\beta(\xi^1)$).

This state of affairs was to be expected.  One knows that for a fundamental extended object, the charge is not localized in space, much as, already for a particle, it is not localized in time along the worldline.  Thus, the only way in which the vector charge $P^\beta_{\textrm{mag}}$ could be contained within the surface is that at each and every point of the worldsheet there should pass a worldline which is tangent to it.  This is what we precisely mean by the term ``interwoven" already used above.

\section{Generalization}
\setcounter{equation}{0}
\label{General}

It is quite straightforward to extend the above treatment to a generalized gauge field corresponding to the Young tableau
$$
 \yng(2,2,2,1,1)
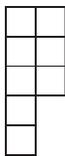
$$
($p$ boxes in the first column, $q$ boxes in the second column with $q \leq p$),  that is, to a source $T^{\al_1 \cdots \al_p \beta_1 \cdots \beta_q}$, with symmetries
\begin{eqnarray}
&&T^{[\al_1 \cdots \al_p] \beta_1 \cdots \beta_q} = T^{\al_1 \cdots \al_p \beta_1 \cdots \beta_q} \label{symm0} \\
&& T^{\al_1 \cdots \al_p [\beta_1 \cdots \beta_q]}= T^{\al_1 \cdots \al_p \beta_1 \cdots \beta_q}  \label{symm1}\\
&& T^{[\al_1 \cdots \al_p \beta_1] \beta_2 \cdots \beta_q}=0. \label{symm2}
\end{eqnarray}
We assume $q \leq D-p-2$.  Otherwise the gauge field carries no degree of freedom since there is no corresponding representation of the little group $SO(D-2)$.  

Now the geometry of the source  consists of a $q$-dimensional flat ``worldline" interwoven into a $p$-dimensional ``worldsheet".  This is because, exactly as in the case considered in the previous section, the $(p+1)$-volume spanned by the $p$-volume of the worldsheet and any tangent vector to the flat worldline should be zero, in order for (\ref{symm2}) to hold.

If the coordinates on the $p$-dimensional worldsheet are denoted by $\xi^0, \cdots, \xi^{p-1}$,  we may take $\xi^0, \cdots, \xi^{q-1}$ as coordinates on the $q$-dimensional interwoven worldline, then the analog of Eq. (\ref{PP2}) is
\be
P^{\beta_1 \cdots \beta_q}   = \frac{m_{\textrm{mag}}}{\sqrt{-\, ^q g}} \frac{\epsilon^{b_1 \cdots b_q}} {q!} \frac{\partial z^{\beta_1}}{\partial \xi^{b_1}} \cdots \frac{\partial z^{\beta_q}}{\partial \xi^{b_q}} \label{GenMom}
\ee
where the indices $b_i$ take the values $0, \cdots, q-1$.  Here, $^q g$ is the determinant of the metric on the interwoven worldsheet. We have been conventional in assuming that both the worldline and the worldsheet have a timelike direction.  However, it has been shown in \cite{Bachas:2009ve} that one can have spacelike currents without any physical inconsistency.  One could thus interweave a spacelike worldline into a timelike worldsheet or one could even take both of them to be spacelike.  In the terminology of \cite{Bachas:2009ve}, we would then be dealing with ``charged events within charged events".  Actually, the consideration of these charged events in the present context is mandatory if one wants to recover the standard results for $D=4$ through compactification from $D=5$.

The generalized strength (\ref{GenMom}) obeys the analog of (\ref{ParTran}), namely
\be
\frac{\partial P^{\beta_1 \cdots \beta_q}_{\textrm{mag}}}{\partial \xi^a} = 0  , \; \; \; \; \; a= 0, \cdots, p-1. \label{ParTran2}
\ee
The solution of (\ref{ParTran2}) that generalizes (\ref{Sol0}) is
\be
\frac{\partial m_{\textrm{mag}}}{\partial \xi^a} = 0
\ee
and
\be
z^\beta (\xi^a) = Z^\beta(\xi^0, \cdots, \xi^{q-1}) + y^\beta(\xi^q, \cdots, \xi^{p})  \label{Sol1}
\ee
with
\be
\frac{\partial^2 Z^\beta}{\partial \xi^a \partial \xi^b} = 0.
\ee

On account of (\ref{ParTran2}), the source  
\begin{eqnarray}
&& \hspace{-.7cm} T^{\alpha_1\cdots  \al_p \beta_1 \cdots \beta_q}(x)   \nonumber \\ 
&& \hspace{-.5cm}  = \int dz^{\alpha_1} \wedge \cdots dz^{\al_p} \delta^{(D)}(x-z(\xi))  P^{\beta_1 \cdots \beta_q}_{\textrm{mag}}   \label{EMTMagGen}
\end{eqnarray}
obeys the conservation law
\be
\partial_{\alpha_1} T^{\alpha_1  \alpha_2 \cdots \alpha_p \beta_1 \cdots \beta_q} = 0.  \label{ConsEMMag2}
\ee

A particular case of the field just described, namely the one where the second column has only one box, describes the magnetic representation of linearized gravitation in an arbitrary space-time dimension $D$.  One has then $p = D-3$, $q=1$.  When there is more than one box in the second column, or the number of boxes in the first column is not equal to $D-3$, one is not dealing with the dual description of gravitation but with a different kind of fields.  Such fields are present in the zero tension limit of string theory.

\section{Quantization Condition}
\setcounter{equation}{0}
\label{Quantization}

To derive the quantization condition between ``electric" and ``magnetic" strengths, one simply notices that the charge $P^{\beta_1 \cdots \beta_q}_{\textrm{mag}}$ appears in place of $q_{\textrm{mag}}$, multiplying the standard conserved current for a $p$-form. Therefore one may fall back into the analysis of \cite{Teitelboim:1985yc} to obtain
\be
\frac{1}{q!} P^{\textrm{el}}_{\beta_1 \cdots \beta_q} P^{\beta_1 \cdots \beta_q}_{\textrm{mag}} = 2 \pi \hbar n \label{QCMSF}
\ee
where $n$ is an integer (see also \cite{Nepomechie:1984wu}).  Here, we have set $8 \pi G=1$ in the electric field equation, as in (\ref{LinEinsEl}).  The  condition (\ref{QCMSF}) generalizes the quantization condition found in \cite{Bunster:2006rt} for gravity in four dimensions.  

Notice that the number of indices on the electric and magnetic charges is the same because to pass from the electric to the magnetic representation one dualizes in the $\alpha$ indices,  which correspond to the first column of the Young tableau.

\section{Concluding Remarks}
\label{Concluding}
\setcounter{equation}{0}

We have shown that the source of a generalized gauge field is a combined extended object: a lower-dimensional current interwoven into a higher-dimensional one.  This construction may be considered as a geometrical realization of the Young tableau describing the symmetries of the field: the whole tableau represents the combined object and its shortest column is the interwoven current.   

The source is extraordinarily rigid since the dynamics of the whole combined current is determined by that of the interwoven one.  This concept cannot be translated straightforwardly into curved spacetime.  But this was to be expected, since generalized gauge fields are known to defy all simple attempts to construct interactions. Indeed, it appears to be necessary to bring in the infinite sequence of fields of all ranks to make them interact consistently \cite{Fradkin:1986qy}. One would expect that the magnetic charge considered herein would emerge as a surface integral at infinity \cite{Regge:1974zd} in the full nonlinear theory.

Our discussion has dealt with generalized gauge fields whose Young tableaux contain two columns. The extension to tableaux with more columns contains the magnetic representation of higher spin fields in higher dimensions.  But the concept of electric-magnetic duality for such fields has not been fully spelled out  at the moment of this writing (see  \cite{Bekaert:2003az} for work in that direction in the sourceless case).  We have hence refrained from making an incursion into that territory.  We plan to address this issue by extending the analysis performed in \cite{Bunster:2006rt} for $D=4$, which includes electric and magnetic sources.

\section*{Acknowledgments} 
Both authors  thank  the Alexander von Humboldt Foundation for Humboldt Research Awards.  The work of M.H. is partially supported by the ERC through the ``SyDuGraM" Advanced Grant, by IISN - Belgium (conventions 4.4511.06 and 4.4514.08) and by the ``Communaut\'e Fran\c{c}aise de Belgique" through the ARC program.  The Centro de Estudios Cient\'{\i}ficos (CECS) is funded by the Chilean Government through the Centers of Excellence Base Financing Program of Conicyt.    The kind hospitality of Hermann Nicolai at the Albert Einstein Institute is gratefuly acknowledged.

\end{document}